\documentclass{elsart}

\begin{document}

\begin{frontmatter}

\title{Black-Scholes model under subordination}

\author{A.~A.~Stanislavsky\corauthref{cor1}}
\corauth[cor1]{E-mail: alexstan@ri.kharkov.ua}

\address{Institute of Radio Astronomy, 4 Chervonopraporna
St., Kharkov 61002, Ukraine}

\begin{abstract}
In this paper we consider a new  mathematical extension of the
Black-Scholes model in which the stochastic time and stock share
price evolution is described by two independent random processes.
The parent process is Brownian, and the directing process is
inverse to the totally skewed, strictly $\alpha$-stable process.
The subordinated process represents the Brownian motion indexed by
an independent, continuous and increasing process. This allows us
to introduce the long-term memory effects in the classical
Black-Scholes model.
\end{abstract}


\end{frontmatter}

The option trading has the long history. The mission of the
options as financial instruments is to protect investors from the
stock market randomness. Since the early seventies the option
market rapidly became very successive in development. The
theoretical study of options was directed in finding a fair and
presumably riskless price of these instruments. Without questions,
the works of Black and Scholes \cite{1} and Merton \cite{2} are a
turning-point in the study. Their method has been proven to be
very useful for investors trading in option markets. On the other
hand, the approach is fruitful for extending the option pricing
theory in many ways. Therefore, nowadays the Black-Scholes (BS)
model is very popular in finance.

The BS equation is nothing else as a diffusion equation. In fact,
their option price formula is a solution of the diffusion equation
with the initial and boundary conditions given by the option
contract terms. The fundamental principles governing the financial
and economical systems are not completely uncovered. In recent
years the physical community has started applying concepts and
methods of statistical and quantum physics of complex systems to
analyze economical problems \cite{3,4,5,6} (and references
therein). The improvement of the BS model itself did not stand
still too. As shown in \cite{7}, the BS equation can be derived
using the Stratonovich calculus. The Gaussian assumption of the
classical BS model based on ideal market conditions simplifies
analytical calculations, but the empirical studies \cite{8} show
the effect of non-ideal market conditions on the true option
price. In particular, the probability distribution of returns has
heavy tails in contrast to a Gaussian. This explains the great
interest to various generalizations of classical results. So, in
\cite{9} the stochastic dynamics of the stock and currency markets
is described by the fractional Langevin-type stochastic
differential equation that differs from the standard Langevin
equation. The continuous-time random walk (CTRW) model is argued
to provide a phenomenological description of tick-by-tick dynamics
in financial markets \cite{10}. The present paper gives arguments
that the CTRW model permits ones to generalize the classical BS
model. This natural extension is based on the general
probabilistic formalism of limit theorems. The important
preference of the approach is its analytical results. We include
the long-term memory effects in the stochastic process of the BS
model. The memory effects are characterized only by one parameter.
To change it, one can control the contribution of memory effects
to the model. The classical BS model is a particular case of the
new model under the complete absence of memory.

The CTRW model is represented by two Markov processes. One of them
corresponds to the random waiting-times between successive jumps,
another defines the random space steps. The geometric Brownian
motion is a special case of the CTRW, where time is deterministic
(see below). Let $T_1$, $T_2$, $\ldots$ be non-negative and
independent identically distributed (i.i.d.) random variables
describing the waiting times between jumps of a walking particle.
Assume that $T_i$ belongs to the strict domain of attraction of
some stable law with index $0<\alpha<1$. This means that there
exist $b_n>0$ ($n\in\mathbf{N}$), and the sum
$b_n(T_1+\cdots+T_n)$ converges in distribution to the process
having the stable distribution with index $\alpha$. The range
$0<\alpha<1$ is conditioned by the support of the time steps $T_i$
on the non-negative semi-axis. In the discrete model the internal
time $\tau$ takes on discrete values with an interval $\delta\tau$
such that $n\leq[\tau/\delta\tau]<n+1$, where $[x]$ denotes the
integer part of $x$. There exists the limit passage from
``discrete steps'' of the CTRW to ``continuous steps''. The
process $b_{[\tau/\delta\tau]} \sum_{i=1}^{[\tau/\delta\tau]}T_i$
under $\delta\tau\to 0$ converges in distribution to a new process
$T(\tau)$ ${d}\atop =$ $\tau^{1/\alpha}T(1)$, where ${d}\atop =$
means equal in distribution, and $T(1)$ ${d}\atop =$ $T_1$. The
new process is Markovian, strictly $\alpha$-stable, totally
skewed. Since $T(\tau)\to\infty$ in probability as
$\tau\to\infty$, the sample paths of $\{ T(\tau)\}$ are increasing
almost surely (a.s.). The process $\{T(\tau)\}$ is self-similar
with exponent $H=1/\alpha>1$ \cite{11}, i.\ e.\ $\{
T(c\tau)\}_{\tau\geq 0}$ ${f.d.}\atop =$ $\{c^{1/\alpha}
T(\tau)\}_{\tau\geq 0}$ for all $c>0$, where ${f.d.}\atop=$
denotes equality of all finite dimensional distributions. Without
loss of generality, we may assume jumps in the one-dimensional
space. Denote by $R_i$ the space steps. Let $R_1$, $R_2$, $\ldots$
be i.i.d. random variables independent of $\{T_i\}$ and have the
Gaussian distribution. Using the limit passage from ``discrete''
to ``continuous'' jumps, we obtain the stochastic process
$\{R(\tau)\}_{\tau\geq 0}$ with the self-similar relation
$\{R(c\tau)\}_{\tau\geq 0}$ ${f.d.}\atop =$
$\{c^{1/2}R(\tau)\}_{\tau\geq 0}$ for any $c>0$. It should be
pointed out that both processes $\{R(\tau)\}_{\tau\geq 0}$ and
$\{T(\tau)\}_{\tau\geq 0}$ depend on the continuous internal
parameter $\tau$ that differs from the real time $t$.

To build the continuous position vector of the walking particle,
we need the process which represents the continuous limit of the
discrete counting process $\{N_t\}_{t\geq 0}$. For $t\geq 0$ the
number of jumps up to time $t$ is $N_t=\max\{n\in\mathbf{N}\mid
\sum_{i=1}^nT_i\leq t\}$, and the vector $\mathbf{r}
_{N_t}=\sum_{i=1}^{N_t}R_i$ defines the position of the particle
at time $t$. It turns out that the scaling limit of
$\{N_t\}_{t\geq 0}$ is the hitting process of $\{T(x)\}_{x \geq
0}$. The hitting time process is well defined $S(t)=\inf\{x\mid
T(x)>t\}$ and depends on the true time $t$. The two processes
$\{T(x)\}$ and $\{S(t)\}$ are the inverse of each other,
$S(T(\tau))=\tau$ a.s. Since $\{T(x)\}_{x\geq 0}$ is strictly
increasing, the process $\{S(t)\}_{t\geq 0}$ is non-decreasing.
From the self-similarity of $\{T(x)\}$ it follows the same
property for $\{S(t)\}$, i.\ e.\ $\{S(ct)\}_{t\geq 0}$
${f.d.}\atop =$ $\{c^\alpha S(t)\}_{t\geq 0}$ for any $c>0$. While
$\{T(x)\}_{x\geq 0}$ is a L\'evy process, the inverse process
$\{S(t)\}_{t\geq 0}$ is no longer a L\'evy process, neither a
Markov process, but it is a continuous submartingal, as shown in
\cite{12}. The random value $S(t)$ has a Mittag-Leffler
distribution with $\langle e^{-vS(t)}\rangle=\sum_{n=0}^\infty
(-vt^\alpha)^n/\Gamma(1+n\alpha)=E_\alpha(-vt^\alpha)$, where
$\langle X\rangle$ denotes the expectation of a real valued random
variable $X$, and $\Gamma(z)$ is the Gamma function. The sample
paths of $\{N_t\}_{t\geq 0}$ and $\{S(t)\}_{t\geq 0}$ are
increasing. Then the position $\mathbf{r}_t$ of the particle at
the given real time $t$ is defined by the subordinated process
$R(S(t))$. Recall briefly that a subordinated process $Y(U(t))$ is
obtained by randomizing the time clock of a random process $Y(t)$
using a new clock $U(t)$, where $U(t)$ is a random process with
nonnegative independent increments. The resulting process
$Y(U(t))$ is said to be subordinated to $Y(t)$, called the parent
process, and is directed by $U(t)$, called the directing process.
The directing process is often referred to as the randomized time
or operational time \cite{13}. In general, the subordinated
process $Y(U(t))$ can become non-Markovian, though its parent
process is Markovian. The process $R(S(t))$ is self-similar with
index $\alpha/2$ such that $\{R(S(ct))\}_{t\geq 0}$ ${f.d.}\atop
=$ $\{c^{\alpha/2}\,R(S(t))\}_{t\geq 0}$ is for all $c>0$. In
fact, the position vector $\mathbf{r}_t=\mathcal{B}_{S(t)}$
represents the randomization of the internal time $\tau$ of a
Brownian motion $\mathcal{B}_\tau$ by an independent, positive and
non-decreasing process $S(t)$.

The probability density of the position vector $\mathbf{r}_t$ with
$t\geq 0$ satisfies
\begin{equation}
p^{\mathbf{r}_t}(t,x)=
\int^\infty_0p^R(\tau,x)\,p^S(t,\tau)\,d\tau, \label{eq1}
\end{equation}
where $p^R(\tau,x)$ represents the probability to find the parent
process $R(\tau)$ at $x$ on the operational time $\tau$, and
$p^S(t,\tau)$ is the probability to be at the operational time
$\tau$ on the real time $t$. The Laplace transform of the
probability density of the random variable $S(t)$ with respect to
$x$ gives
\begin{displaymath}
\bar{p}^S(t,v)=\int^\infty_0e^{-vx}\,p^S(t,x)\,dx=\langle
e^{-vS(t)}\rangle=E_\alpha(-vt^\alpha).
\end{displaymath}
We need also the Laplace transform of $p^S(t,x)$ with respect to
$t$. The Mittag-Leffler function $E_\alpha(-vt^\alpha)$ has the
following Laplace transform $u^{\alpha-1}/(u^\alpha+v)$ with
respect to $t$. To invert the latter analytically, we obtain
\begin{displaymath}
\hat{p}^S(u,x)=\int^\infty_0e^{-ut}\,p^S(t,x)\,dt=u^{\alpha-1}
\exp\{-u^\alpha x\}.
\end{displaymath}
In Laplace space the probability density $p^{\mathbf{r}_t}(t,x)$
has the most simple form $u^{\alpha-1}\hat{p}^R(u^\alpha,x)$, as
$\hat{p}^R(u^\alpha, x)=\int_0^\infty
p^R(\tau,x)\,\exp\{-u^\alpha\tau\}\,d\tau$. For our purpose, it is
useful to find the explicit form of the probability density
$p^{\mathbf{r}_t}(t,x)$. According to the inverse formula applied
to $\hat{p}^S(u,x)$, we have
\begin{equation}
p^{S}(t,x)=\frac{1}{2\pi
i}\int_{Br}e^{ut-xu^\alpha}\,u^{\alpha-1}\,du\,, \label{eq2}
\end{equation}
where $Br$ denotes the Bromwich path. Make the variable transform
$ut\to u$ and denote $w=x/t^\alpha$. Then we deform the Bromwich
path into the Hankel path $Ha$ for which a contour begins at
$u=-\infty-ia$ $(a>0)$, encircles the branch cut that lies along
the negative real axis and comes to the end at $u=-\infty+ib$
$(b>0)$. Expanding function $\exp\{-wu^\alpha\}$ in a Taylor
series about $w$ and using the Hankel representation of the
reciprocal of the Gamma function
\begin{displaymath}
\frac{1}{\Gamma(z)}=\frac{1}{2\pi i}\int_{Ha}e^u\,u^{-z}\,du,
\end{displaymath}
we get the following series
\begin{displaymath}
p^S(t,x)=t^{-\alpha}\sum_{k=0}^\infty\frac{(-x/t^\alpha)^k}{k!
\Gamma(1-\alpha-k\alpha)}=t^{-\alpha}F_\alpha(x/t^\alpha).
\end{displaymath}
Further, we briefly consider the character of $F_\alpha(z)$.

The function $F_\alpha(z)$ is an entire function in $z$. It has
the H-function representation $H^{10}_{11}\left(z\mid{(1-\alpha,
\alpha)\atop (0,1)}\right)$ \cite{14}. The important property of
$F_\alpha(z)$ is that it is non-negative for $z>0$. It is easily
verified that $\int_0^\infty F_\alpha(z)\,dz=1$.  Thus, the
function can be a probability density. The case $\alpha=1$
corresponds to the Dirac $\delta$-function, $F_1(z)=\delta(z-1)$.
In particular cases $\alpha=1/2$ and $\alpha=1/3$ we have
$F_{1/2}(z)=\exp\{-z^2/4\}/\sqrt{\pi}$ and $F_{1/3}(z)=3^{2/3}{\rm
Ai}(z/3^{1/3})$ respectively, where $Ai$ denotes the Airy function
\cite{15}. The function $F_\alpha(z)$ has also other interesting
properties. For $0<\alpha\leq 1/2$ the function is monotonic
decreasing, whereas for $1/2<\alpha<1$ it has a maximum value at a
certain point $z_{\rm max}$ depending on $\alpha$. It should be
observed here that the basic Cauchy and Signaling problems of the
time fractional diffusion-wave equation can be expressed in terms
of the function $F_\alpha(z)$ \cite{16,17}.

Turning back to Eq. (\ref{eq1}), the probability density
$p^{\mathbf{r}_t}(t,x)$ is written as
\begin{equation}
p^{\mathbf{r}_t}(t,x)= \int^\infty_0F_\alpha(z)\,p^R(t^\alpha z
,x)\,dz=\frac{1}{\sqrt{\pi Dt^\alpha}
}\int^\infty_0F_\alpha(z)\,e^{-x^2/(Dt^\alpha z
)}\,\frac{dz}{\sqrt{z}}\,, \label{eq3}
\end{equation}
where $D$ is the constant. This function is non-negative and
satisfies the normalization condition
\begin{displaymath}
\int^\infty_{-\infty}p^{\mathbf{r}_t}(t,x)\,dx=\int_0^\infty
F_\alpha(z)\,dz=1.
\end{displaymath}
Since the parent process $\{R(\tau)\}$ and the directing process
$\{S(t)\}$ have finite moments of any order, the subordinated
process $\{R(S(t))\}$ has finite moments of any order too. The
first and second moments of $\mathbf{r}_t$ can be obtained by the
direct calculations:
\begin{eqnarray}
\langle\mathbf{r}_t\rangle&=&0\,,\nonumber\\
\langle\mathbf{r}_t^2\rangle&=&\frac{1}{2}\,Dt^\alpha\int^\infty_0z\,
F_\alpha(z)\,dz=\frac{Dt^\alpha}{2\Gamma(1+\alpha)}\,. \nonumber
\end{eqnarray}
The process $\mathbf{r}_t$ behaves as subdiffusion ($0<\alpha<1$).
Note that the boundary case $\alpha=1$ may be also included in the
consideration because of $T(\tau)=\tau$ a.s. Then the hitting time
process is deterministic, $S(t)=t$. The probability density
$p^S(\tau,t)$ degenerates in the Dirac $\delta$-function so that
$p^{\mathbf{r}_t}(t,x)$ becomes equal to $p^R(t,x)$.  The constant
$D$ is interpreted as a generalized diffusion coefficient with
dimension $[D]={\rm length}^2/\,{\rm time}^\alpha$.

The ordinary Brownian motion satisfies the stochastic differential
equation (SDE)
\begin{displaymath}
dR(\tau)=f(R(\tau))\,d\tau+g(R(\tau))\,d\mathcal{B}_\tau\,,
\end{displaymath}
where $f$ and $g$ are some functions. The process subordinated to
the Brownian motion $\{R(S(t))\}_{t\geq 0}=\mathcal{B}_{S(t)}$ is
a continuous martingal and the directing process $\{S(t)\}$ is a
continuous submartingal with respect to an appropriate filtration
\cite{18}. Therefore, the subordinated process obeys the following
SDE
\begin{displaymath}
d\mathbf{r}_t=f(\mathbf{r}_t)\,dS(t)+g(\mathbf{r}_t)\,
d\mathcal{B}_{S(t)}\,.
\end{displaymath}
In the classical BS model the evolution of the option price is
governed by the Brownian motion. The well-known BS formula is of
the form
\begin{displaymath}
\mathcal{C}(\tau,x)=x\Phi(d_+)-Ke^{-\beta\tau}\Phi(d_-)\,,\quad
\beta=2r/\sigma^2\,,
\end{displaymath}
where $x$ is the share price, $K$ the striking price, $r$ the
interest rate, $\sigma$ the volatility, and the probability
integral
\begin{displaymath}
\Phi(z)=\frac{1}{\sqrt{2\pi}}\int_{-\infty}^{z}\exp\{-y^2/2\}\,dy
\end{displaymath}
is calculated for
\begin{displaymath}
d_\pm=(2\tau)^{-1/2}\Bigr[\ln(x/K)+\tau(\beta\pm 1)\Bigl].
\end{displaymath}
If the price evolution is consequent of the subordinated process
$\mathcal{B}_{S(t)}$, the BS formula transforms into
\begin{equation}
\mathcal{S}(t,x)=t^{-\alpha}\int^\infty_0F_\alpha(z/t^\alpha)\,
\mathcal{C}(z,x)\,dz\,.\label{eq4}
\end{equation}
At $\alpha=1$ we obtain the classical BS formula.  All financial
derivatives (options of any kind, futures, forwards, etc.) have
the same boundary conditions, but different either initial or
final condition \cite{19}. The detailed comparison of the various
cases for this new model (\ref{eq4}) will carry out elsewhere. The
fractional extension of the BS model has been considered also in
\cite{20}, but on the macroscopic  basis without any microscopic
dynamics presented above.

Finally, we note that the index $\alpha$ characterizes memory
effects in the subordinated process $\mathbf{r}_t$. Let $L(x)$ be
a time-independent Fokker-Plank operator, whose exact form is not
important for the following. If the ordinary Fokker-Plank equation
(FPE) $\partial p^R(\tau,x)/\partial\tau=[L(x)\,p^R](\tau,x)$
describes the evolution of a Brownian particle, the probability
density $p^{\mathbf{r}_t}(t,x)$ satisfies the fractional FPE. This
can be shown by simple computations. Using the relation
$\hat{p}^{\mathbf{r}_t}(u,x)=u^{\alpha-1} \hat{p}^R(u^\alpha,x)$
in Laplace space and acting the operator $L(x)$ on
$\hat{p}^{\mathbf{r}_t}(u,x)$, the Laplace image
$[L(x)\,\hat{p}^{\mathbf{r}_t}](u,x)$ takes the form
$u^\alpha\,\hat{p}^{\mathbf{r}_t}(u,x)-f(x)\,u^{\alpha-1}$, where
$f(x)$ is the initial condition. The inverse Laplace transform of
the latter expression gives the above-mentioned fractional FPE
\begin{displaymath}
p^{\mathbf{r}_t}(t,x)=f(x)+\frac{1}{\Gamma(\alpha)}\int_0^td\tau
(t-\tau)^{\alpha-1}[L(x)\,p^{\mathbf{r}_t}](\tau,x)\,.
\end{displaymath}
The kernel of this integral equation is a power function. It just
causes the long-term memory effects in the process of interest. As
shown in \cite{17}, due to such kind of memory effects, the
complex nature of the microscopic behavior of stochastic systems
can be transmitted to the macroscopic level of their dynamics.


\end{document}